      [1999/12/01 v1.4c Il Nuovo Cimento]
\documentclass[final]{cimento}

\usepackage{graphicx}  
\title{PROMPT\ETC: Panchromatic Robotic Optical Monitoring and Polarimetry Telescopes}
\author{D.~Reichart\from{ins:x}\ETC,
M.~Nysewander\from{ins:x},
J.~Moran\from{ins:x},
J.~Bartelme\from{ins:x},
M.~Bayliss\from{ins:x},
A.~Foster\from{ins:x},
J.~C.~Clemens\from{ins:x},
P.~Price\from{ins:y},
C.~Evans\from{ins:x},
J.~Salmonson\from{ins:z},
S.~Trammell\from{ins:a},
B.~Carney\from{ins:x},
J.~Keohane\from{ins:b}
        \atque
R.~Gotwals\from{ins:c}}
\instlist{\inst{ins:x} Department of Physics and Astronomy, University of North Carolina at Chapel Hill, Chapel Hill, NC
  \inst{ins:y} Institute for Astronomy, University of Hawaii, Honolulu, HI
  \inst{ins:z} Lawrence Livermore National Laboratory, University of California, Livermore, CA
  \inst{ins:a} Department of Physics and Optical Sciences, University of North Carolina at Charlotte, Charlotte, NC
  \inst{ins:b} Department of Physics and Astronomy, Hampden-Sydney College, Hampden-Sydney, VA
  \inst{ins:c} Morehead Planetarium and Science Center, University of North Carolina at Chapel Hill, Chapel Hill, NC}
\PACSes{\PACSit{95.45.Cs}{Ground-based ultraviolet, optical and infrared telescopes}
\PACSit{95.75.Hi}{Polarimetry}
\PACSit{95.75.Rs}{Remote observing techniques}
\PACSit{98.70.Rz}{Gamma-ray sources; gamma-ray bursts}
\PACSit{98.80.Es}{Observational cosmology}}
\begin{document}

\maketitle

\begin{abstract}
Funded by \$1.2M in grants and donations, we are now building PROMPT at CTIO.  When completed in late 2005, PROMPT will consist of six 0.41-meter diameter Ritchey-Chr\'etien telescopes on rapidly slewing mounts that respond to GRB alerts within seconds, when the afterglow is potentially extremely bright.  Each mirror and camera coating is being optimized for a different wavelength range and function, including a NIR imager, two red-optimized imagers, a blue-optimized imager, an UV-optimized imager, and an optical polarimeter.  PROMPT will be able to identify high-redshift events by dropout and distinguish these events from the similar signatures of extinction.  In this way, PROMPT will act as a distance-finder scope for spectroscopic follow up on the larger 4.1-meter diameter SOAR telescope, which is also located at CTIO.  When not chasing GRBs, PROMPT serves broader educational objectives across the state of North Carolina.  Enclosure construction and the first two telescopes are now complete and functioning:  PROMPT observed Swift's first GRB in December 2004.  We upgrade from two to four telescope in February 2005 and from four to six telescopes in mid-2005.  
\end{abstract}

\section{PROMPT}
PROMPT (Panchromatic Robotic Optical Monitoring and Polarimetry Telescopes) is a robotic telescope system that will provide well sampled multiwavelength light curves and polarization histories of GRB afterglows beginning only seconds after spacecraft notification.  When completed in late 2005, PROMPT will consist of six 0.41-meter Ritchey-Chr\'etien telescopes by RC Optical Systems on rapidly slewing Paramount ME mounts by Software Bisque.  Each mirror and camera coating is being optimized for a different wavelength range and function, including a NIR imager, two red-optimized imagers, a blue-optimized imager, an UV-optimized imager, and an optical polarimeter.  The NIR imager is a LN2-cooled MicroCam by Rockwell Scientific, which uses a PICNIC 256 $\times$ 256 HgCdTe FPA (same as NICMOS3) and will have a 5' field of view and 1.2" pixels.  The optical imagers are Alta U47$+$s by Apogee, which use back-illuminated E2V 1024 $\times$ 1024 CCDs with $\approx$1-second readout times and will have 10' fields of view and 0.6" pixels.  The optical polarimeter consists of a Fresnel rhomb $+$ Wollaston prism assembly that we are designing and building at the Abraham Goodman Laboratory for Astronomical Instrumentation at the University of North Carolina at Chapel Hill.  It will produce two orthogonally polarized 5' fields of view on another Alta U47+.  Stokes parameters for linear polarization are measured from the ratio of the images for different rotation angles of the Fresnel rhomb.

PROMPT's six imager design (counting the polarimeter) will allow the
u'g'r'R$_{\rm c}$i'z'JH SFD of the afterglow to be
reconstructed, modulo a small, calibration-induced slope when non-photometric,
after only two, mostly redundant exposure sets -- regardless of potentially
great variations in the early-time light curve -- simply by repeating
(preferably long-wavelength) filters across consecutive exposure sets:  With
only two cameras at least seven exposure sets would be needed, and with only one
camera this would not be possible until after the afterglow settled into an
easy-to-model behavior.  From the reconstructed SFD the redshift of the GRB
can be estimated from dropout, and because of the eight-band coverage the
dropout signature can be distinguished from the similar but broader signatures
of extinction by dust, both in our galaxy and in the host galaxy, using the
extinction-curve, Ly$\alpha$-forest, and Lyman-limit models of~\cite{ref:r01}.
Since the possibility of an early-time color transition might cause confusion,
the technique will be repeated to confirm the result.  

PROMPT's polarimeter will also forge new ground.  When we first proposed to
build a robotic polarimeter, polarizations of only a few percent had been
measured, and then only at intermediate and late times.
However, since then the solar gamma-ray spacecraft RHESSI's
serendipitous detection of GRB 021206 resulted in a 
polarization measurement for the GRB of (80 $\pm$ 20)\% (\cite{ref:cb03};
however, see~\cite{ref:rf04}), which suggests that the jet that produces
the GRB, at least initially, might not be completely hydrodynamically driven,
but rather might have a magnetically driven component through an almost completely
ordered field.  If this is indeed the case, the reverse shock, which is most
likely responsible for the prompt optical emission, would propagate back
through this highly ordered field and consequently the prompt emission would
be similarly polarized (e.g.,~\cite{ref:gk03}).  Hence, the optical
polarization might actually be highest, and potentially extremely so, right
when the afterglow is brightest, and potentially extremely so -- and
consequently almost trivial to measure if one has a robotic polarimeter.  When
completed, PROMPT will be able to measure
optical polarizations beginning only $\approx$20 seconds after the burst, while
many
of the long-duration GRBs are still bursting.  

The construction of PROMPT's first four telescopes is well timed with the beginning of operations for Swift.
Swift will revolutionize the GRB
field for the second time in eight years:  Swift is expected to localize $\sim$100 GRBs per year with localizations accurate to 1' -- 4' relayed to
observers on
the ground within $\approx$15 seconds of the burst and localizations
accurate to
many arcseconds relayed within minutes of the burst.  Swift will also localize
relatively unextinguished, $z < 3 - 4$ GRBs (probably about half of their
long-duration/soft-spectrum GRBs) to a few tenths of an arcsecond on a longer
timescale with UVOT.  The expected rate of GRBs that will be observable from
the PROMPT site (i.e., that occur sufficiently above the horizon when the sun
is below the horizon and the weather is acceptable) during the Swift era (including HETE-2, Integral, IPN, etc.) is
$\sim$15 per year that will be observable within minutes of the burst and another
$\sim$30 per year that will be observable on longer timescales as the earth rotates
the GRB field above the horizon and/or the sun below the horizon.  Of the $\sim$15
GRBs per year that PROMPT will observe on the rapid timescale, $\sim$2 or 3 are expected
to be at redshifts greater than 5 and $\sim$1 is expected to be more distant
than the most distant object in the universe yet identified, based on various
best guesses about the star-formation rate at high redshifts (e.g.,~\cite{ref:lr00};~\cite{ref:cl00};~\cite{ref:bl02}).  Over
an extended mission, we expect to detect and identify GRBs with redshifts as
high as $\sim$10 -- 15, reaching back to when some of the first stars in the
universe formed.

\begin{figure}
\includegraphics[height=.293\textheight]{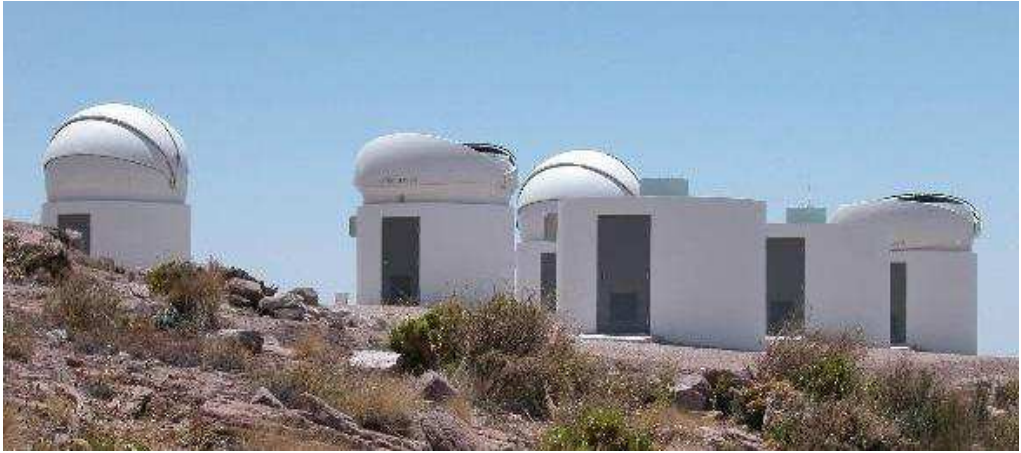}     
\caption{The six PROMPT enclosures at CTIO as of January 15, 2005. Photo by Enrique Figueroa.}
\end{figure}

\begin{figure}
\includegraphics[height=.35\textheight]{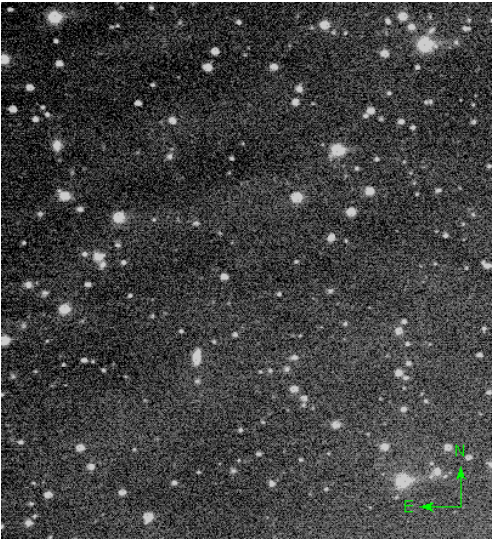}     
\caption{PROMPT's first science image, of Swift's first GRB (041217).  No obvious transients were found to a 3-sigma limiting magnitude of R$_{\rm c} = 21.5$ mag at a mean time of 23.8 hours after the burst~\cite{ref:bea04}.}
\end{figure}

\section{SOAR}
By locating PROMPT on Cerro Tololo near the 4.1-meter diameter SOAR telescope (which is less expensive and more
convenient than locating PROMPT on Cerro Pachon next to SOAR), we ensure that
every GRB that is observable to PROMPT is also observable to SOAR.  Given
SOAR's rapid response capabilities -- rapid slewing and instrument changing, an interrupt policy that was designed with GRBs in mind and spans partner institutions, and remote observing centers -- this ensures that a
spectroscopic redshift will also be measured for nearly every GRB that PROMPT detects, even
if faint or rapidly fading.  Although we do not plan to use PROMPT for target
selection for SOAR, PROMPT will prove useful for deciding which spectrograph
to use once SOAR is in position:  If PROMPT finds a GRB to be at high redshift
or highly extinguished, we will stick with our default, NIR spectrograph
(OSIRIS, resolution = 3,000).  Otherwise, we will switch to the optical
spectrograph (UNC-Chapel Hill's Goodman Spectrograph, resolution = 8,000), a
change that we can make in under a minute.  Swift's UVOT, although unable to
distinguish between high-redshift and highly extinguished GRBs, will provide
similar information, but on a longer timescale.    


In addition to using SOAR to complement PROMPT's u'g'r'R$_{\rm c}$i'z'JH spectral coverage with NIR and optical spectroscopy, we will take a few K-band images with SOAR to extend PROMPT's redshift range from $z \approx 13$ to $z \approx 17$.  This spectral coverage synergizes nicely with Swift's UVOT's ultraviolet and UBV spectral coverage, not to mention PROMPT's typically half-minute faster response time.  

\acknowledgments
DER very gratefully acknowledges support from NSF's MRI and PREST programs,
NASA's APRA, Swift GI and IDEAS programs, Dudley Observatory's Ernest F. Fullam Award, and especially Leonard Goodman and Henry Cox.

\end{document}